\documentstyle[12pt,epsfig]{article}
\newcommand{\beq}{\begin{equation}}
\newcommand{\eeq}{\end{equation}}
\newcommand{\psibar}{\overline{\psi}}
\newcommand{\la}{\langle}
\newcommand{\ra}{\rangle}
\newcommand{\amp}[1]{\la #1 \ra}

\newcommand{\ba}{\begin{eqnarray}}
\newcommand{\ea}{\end{eqnarray}}
\newcommand{\dz}{\int \frac{d^{4}z}{(2\pi)^4}}

\newcommand{\slsh}[1]{\mbox{$\not\! #1$}}
\newcommand{\bm}[1]{\mbox{\boldmath $#1$}}
\newcommand{\text}[1]{{\rm #1}}

\begin{document}

\begin{flushright}
\begin{minipage}{3.5 cm}
{\small hep-ph/9710525\\
NIKHEF 97-041\\ 
VUTH 97-17}
\end{minipage}
\end{flushright}
\begin{center}
{\LARGE\bf Single spin asymmetries in the Drell-Yan
process\footnote{Invited talk presented at the workshop 
'Deep inelastic scattering off polarized targets: theory meets experiment', 
DESY-Zeuthen, September 1-5, 1997}}

\vspace{1cm}
{\underline{D. Boer}$^a$, P.J. Mulders$^{a,b}$, O.V. Teryaev$^c$}

\vspace*{1cm}
{\it $^a$National Institute for Nuclear Physics and High--Energy
Physics (NIKHEF), P.O. Box 41882, NL-1009 DB Amsterdam, the Netherlands}\\

\vspace*{3mm}
{\it $^b$Department of Physics and Astronomy, Free University, 
De Boelelaan 1081, NL-1081 HV Amsterdam, the Netherlands}\\

\vspace*{3mm}
{\it $^c$Joint Institute for Nuclear Research, 141980 Dubna, Russia}\\

\vspace*{2cm}

\end{center}

\begin{abstract}
We discuss single transverse spin asymmetries in the Drell-Yan process 
originating from 
so-called gluonic poles in twist-three 
hadronic matrix elements, as first considered by Qiu and Sterman.
Even though time-reversal invariance is not broken, 
the effects of such poles cannot be distinguished from those of 
time-reversal odd distribution functions.
We show the connection between gluonic poles and  
large distance gluon fields, in particular we focus on boundary conditions. 
We identify the possible single spin asymmetries in the Drell-Yan process.  
\end{abstract}

\section{Introduction}

\vspace{1mm}
\noindent
In the standard description of the Drell-Yan process (DY) in terms of 
distribution functions time-reversal symmetry implies the absence of
single spin asymmetries 
at tree level, even including order $1/Q$ 
corrections \cite{Tangerman-Mulders-95a}.
Additional time-reversal odd (T-odd) 
distribution functions (DFs) are 
present when the incoming hadrons cannot be treated as plane-wave states.
This may occur due to some factorization breaking mechanism \cite{Anselmino}. 
We will show that, even apart from such mechanisms, the 
contributions of T-odd DFs may effectively arise due 
to the presence of so-called gluonic poles attributed to large distance 
gluon fields. 
The gluonic poles appearing in the 
twist-three hadronic matrix elements 
\cite{QS-91b,Korotkiyan-T-94,E-Korotkiyan-T-95} together with 
imaginary phases of hard 
subprocesses effectively give rise to the same single spin asymmetries 
as T-odd DFs, but without a violation of time-reversal invariance.
This is the origin of the single spin asymmetry of Ref.\ \cite{Hammon-97}. 
For a detailed account on these matters see \cite{BMT-97}.

\section{The DY process in terms of distribution functions}
\vspace{1mm}
\noindent

We employ methods originating from Refs.\ 
\cite{Soper-77,Ralst-S-79,Coll-S-82,Politzer-80,EFP-83,Efremov-Teryaev-84,Jaffe-Ji-91} 
in order
to describe the soft (non-perturbative) parts of the scattering process in 
terms of correlation functions, which are (Fourier transforms of) hadronic
matrix elements of non-local operators. We restrict ourselves to tree-level, 
but 
include
$1/Q$ power corrections. The asymmetries under investigation are loosely
referred to as 'twist-three' asymmetries, since they are suppressed by a 
factor 
of $1/Q$, where the photon
momentum $q$ sets the scale $Q$, such that $Q^2=q^2$. We do not take $Z$
bosons into account, since the asymmetries are likely to be negligible at or
above the $Z$ threshold. 

\begin{figure}[htb]
\begin{center}
\leavevmode \epsfxsize=6.5cm \epsfbox{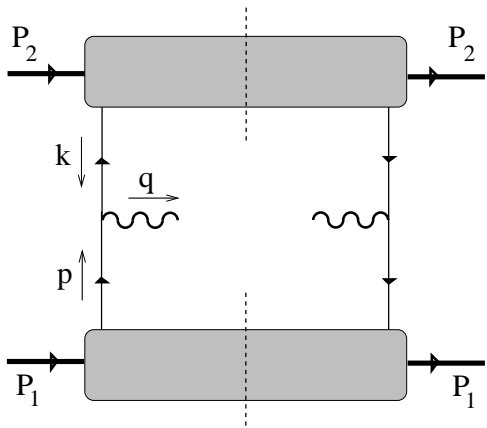}\\
\vspace{0.2 cm}
\caption{\label{LODY}The leading order contribution to the
Drell-Yan process}
\end{center}
\end{figure}

The Drell-Yan process consists of two soft parts (depicted in Fig.\ \ref{LODY}
for the leading order) 
and one of them is described
(up to order $1/Q$) by the quark correlation functions $\Phi$ and $\Phi_A$, 
and the other soft part by the
antiquark correlation functions, denoted by $\overline \Phi$ and 
$\overline \Phi{}_A^\alpha$. 
The quark-quark correlation function,
\beq
\Phi_{ij}(P_1,S_1;p)=\dz e^{ip \cdot z} 
\amp{P_1,S_1|\psibar_j (0) \psi_i (z) | P_1,S_1},
\eeq
is a function of 
the momentum and spin vectors $P_1, S_1$ of the incoming hadron (spin-1/2),
with $P_1\cdot S_1=0$,  
and the quark momentum $p$. The 
hadron momentum $P_1$ is chosen to be predominantly along a light-like
direction given by the vector $n_+$. Another light-like direction $n_-$ is 
chosen such that 
$n_+\cdot n_-=1$; both vectors are dimensionless. The second hadron is chosen 
to be predominantly in the $n_-$ direction, such that $P_1 \cdot P_2 = {\cal
O} (Q^2)$. 
We write $p^\pm=p\cdot n_\mp$ and 
approximate the parton momentum $p \approx  x P_1 + p_T$ and the polarization 
vector $S_1 \approx  \lambda_1P_1/M_1  + S_{1T}^{}$. 

We will consider the case where one integrates over the transverse 
momentum $q_T$ of the photon. One then only encounters 
correlation functions integrated over all but the leading component, 
such that they are functions of the light-cone
momentum fractions (e.g.\ $x$) only. 
So we consider the partly integrated quark 
correlation functions
\ba
&& \Phi_{ij} (x) \equiv  \int \frac{d \lambda}{2\pi}e^{i\lambda x}\langle \,
P_1,S_1|\overline{\psi}_j (0) \psi_i(\lambda n_-)| P_1,S_1 \, \rangle,\\[2 mm]
&& \Phi_{Aij}^{\alpha} (x,y) \equiv \int \frac{d \lambda}{2\pi} 
\frac{d \eta}{2\pi} e^{i\lambda x} e^{i\eta (y-x)} 
\amp{P_1,S_1|\psibar_j (0) g A _T^{\alpha}(\eta n_-) 
\psi_i(\lambda n_-)| P_1,S_1}.
\ea
The anti-quark correlation function is defined as
\ba
&& \overline \Phi(\bar x)=  
\int \frac{d \lambda}{2\pi} e^{-i\lambda \bar x} 
\langle P_2,S_2 \vert \psi(\lambda n_+) \overline \psi(0)
\vert P_2,S_2 \rangle
\ea
and the function $\overline \Phi_A^\alpha(\bar x,\bar y)$ 
is defined analogously. 
Moreover, the inclusion of path-ordered exponentials, such as,
\beq
{\cal L}(0,\lambda n_-) = {\cal P} \exp \left(-ig\int_0^{\lambda n_-} 
dz^\mu \,A_\mu(z)\right),
\eeq
which are needed in 
order to render the correlation functions gauge invariant, is implicit.

\begin{figure}[htb]
\begin{center}
\leavevmode \epsfxsize=6.5cm \epsfbox{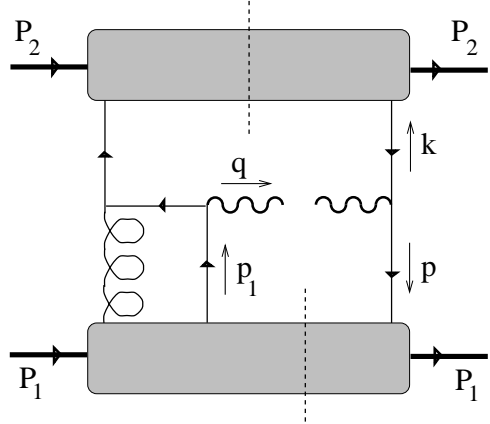}
\vspace{0.2 cm}
\caption{\label{SLOfig}Subleading order contribution to the
Drell-Yan process}
\end{center}
\end{figure}

In the expression of the hadron tensor the fermion propagators appearing 
in the hard 
part of the subleading contributions (cf.\ Fig.\ \ref{SLOfig}) 
is approximated like (neglecting contributions that will appear 
suppressed by $1/Q^2$)  
\begin{eqnarray}
\frac{\slsh{p_1}-\slsh{q}+m}{(p_1-q)^2-m^2+ i \epsilon} &\approx& 
-\frac{\slsh{n_+}}{Q\sqrt{2}}\, \frac{x-y}{x-y + i \epsilon}.
\label{prop}
\end{eqnarray}
Hence, a zero-momentum gluon ($x=y$) is always accompanied by an 
on-shell quark propagator, where we note the following:
\beq
\int\, dy \, \Phi_A^\alpha (x,y)  \, \frac{x-y}{x-y + i\epsilon}
\stackrel{\Phi_A^\alpha(x,x)=0}{\Longrightarrow} 
\int\, dy \, \Phi_A^\alpha (x,y).
\eeq

For the correlation functions $\Phi$ and $\Phi^{\alpha}_{A}$ we need up to 
order $1/Q$ 
the following parametrizations 
in terms of distribution functions~\cite{Jaffe-Ji-91}:
\begin{eqnarray}
\Phi(x)&=&\frac{1}{2} \left[f_1(x) \mbox{$\not\! P_1\,$} + 
g_1(x)\,\lambda_1\gamma_{5}\mbox{$\not\! P_1\,$}
+ h_1(x)\,\gamma_{5}\mbox{$\not\! S$}_{1T}\mbox{$\not\! P_1\,$}
\right] \nonumber\\[3 mm]
&+& \frac{M_1}{2} 
\left[ e(x)\bm{1} + g_T(x) \gamma_5 \mbox{$\not\! S$}_{1T}
+ h_L(x) \frac{\lambda_1}{2} \gamma_5 \left[\slsh{n_+}, \slsh{n_-} \right]
\right],\label{paramPhi}\\[3 mm]
\Phi^{\alpha}_{A}(x,y)&=& 
\frac{M_1}{2} \bigg[ G_A(x,y)\,i\epsilon_T^{\alpha\beta} 
S_{1T \, \beta} \mbox{$\not\! P_1\,$} + \tilde{G}_A(x,y)\,
S_{1T}^\alpha \gamma_{5} \mbox{$\not\! P_1\,$}  \nonumber \\[3 mm]
&+& H_A(x,y) \lambda_1\gamma_{5} \gamma_T^\alpha \mbox{$\not\! P_1\,$}
+E_A(x,y)  \gamma_T^\alpha \mbox{$\not\! P_1\,$} \bigg],
\label{paramPhiA}
\ea
where $\epsilon_T^{\mu\nu}=\epsilon^{\alpha \beta\mu\nu} n_{+\alpha}
n_{-\beta}$.

The parametrization of $\Phi(x)$ is consistent with requirements 
following from hermiticity, parity and time-reversal invariance, 
\begin{eqnarray}
\Phi^\dagger (P_1,S_1;p) = \gamma_0 \,\Phi(P_1,S_1;p)\,\gamma_0 
& & \quad \mbox{[Hermiticity]} \\
\Phi(P_1,S_1;p) = \gamma_0 \,\Phi(\bar P_1,-\bar S_1;\bar p)\,
\gamma_0 & & \quad \mbox{[Parity]} \\
\Phi^\ast(P_1,S_1;p) = \gamma_5 C \,\Phi(\bar P_1,\bar S_1;\bar p)\, 
C^\dagger \gamma_5 & & \quad \mbox{[Time\ reversal]} 
\label{Tinvariance}
\end{eqnarray}
where 
$\bar p$ = $(p^0,-\bm{p})$, etc. 
For the one-argument functions in Eq.\ (\ref{paramPhi}) it follows from 
hermiticity 
that they are real. Note 
that for the validity of Eq.\
(\ref{Tinvariance}) it is essential that the incoming hadron is a plane wave
state. For $\Phi_A^\alpha$  
hermiticity, parity and time-reversal invariance yield the following 
relations:
\begin{eqnarray}
\left[\Phi_A^\alpha(P_1,S_1;p_1,p_2)\right]^\dagger = 
\gamma_0 \,\Phi_A^\alpha(P_1,S_1;p_2,p_1)\,\gamma_0
& &\quad \mbox{[Hermiticity]} \\
\Phi_A^\alpha(P_1,S_1;p_1,p_2) = \gamma_0 
\,\Phi_{A\alpha}(\bar P_1,-\bar S_1;\bar p_1, \bar p_2 )\,
\gamma_0 & & \quad \mbox{[Parity]} \\
\left[\Phi_A^\alpha(P_1,S_1;p_1,p_2)\right]^\ast = 
\gamma_5 C \,\Phi_{A\alpha}(\bar P_1,\bar S_1;\bar p_1, \bar p_2  )\, 
C^\dagger \gamma_5 & &\quad \mbox{[Time\ reversal]} 
\label{TinvA}
\end{eqnarray}
Hermiticity then gives for the two-argument functions in Eq.\ 
(\ref{paramPhiA}) the following constraints:
\ba
G_A(x,y)= -G_A^\ast (y,x),&& 
\quad \tilde{G}_A(x,y) =\tilde{G}_A^\ast (y,x),\\
E_A(x,y)= -E_A^\ast (y,x), && 
\quad H_A(x,y)= H_A^\ast (y,x).
\ea
Hence, the real and imaginary parts of these two-argument functions have
definite symmetry properties under the interchange of the two arguments. 
If we would impose time-reversal invariance all four functions must be real
and $\tilde{G}_A$ and $H_A$ are then symmetric and 
$G_A$ and $E_A$ are antisymmetric
under interchange of the two arguments, such that at $x=y$ only $\tilde{G}_A$ 
and $H_A$ survive.

In the remainder of this section we do not impose time-reversal invariance 
and hence allow for 
imaginary parts of these functions. In addition, the following 
(T-odd) one-argument
DFs then appear:
\begin{eqnarray}
\left. \Phi(x) \right|_{\text{T-odd}}&=&\frac{M_1}{2} \left[ f_T(x)
\epsilon_T^{\mu\nu} S_{1T\mu}\gamma_{T\nu} - e_L(x) \lambda_1 i\gamma_5
+ h(x) \frac{i}{2} \left[\slsh{n_+},\slsh{n_-} \right]
\right]. 
\ea 

The two-argument functions and the one-argument functions are related by 
the classical e.o.m.\ ($(i\slsh{\! D}-m) \psi=0$), 
which hold inside hadronic matrix elements \cite{Politzer-80}. 
Using a similar parametrization for $\Phi_D^\alpha$ (defined like 
$\Phi_A^\alpha$, but with $gA_T^\alpha$ replaced by $iD_T^\alpha$) 
as in Eq.\ (\ref{paramPhiA}),
one has the following relations \cite{Efremov-Teryaev-84,Jaffe-Ji-91}: 
\ba
& & \int dy \, 
\Bigl[{\rm Re} \, G_D(x,y) + {\rm Re} \, \tilde{G}_D(x,y) \Bigr] 
=2x g_T(x) - 2\frac{m}{M}h_1(x),
\label{eom}\\
& & \int dy \, 
\Bigl[{\rm Im} \, G_D(x,y) + {\rm Im} \, \tilde{G}_D(x,y) \Bigr] = 2i x f_T(x),
\label{Imeom1}\\
& & \int dy \,  \Bigl[2 \, {\rm Re} \, H_D(x,y) \Bigr] = 
x h_L(x) -\frac{m}{M} g_1(x),\\
& & \int dy \,  \Bigl[2 \, {\rm Im} \, H_D(x,y) \Bigr] =
-i x e_L (x),
\label{Imeom2}\\
& & \int dy \,  \Bigl[2 \, {\rm Re} \, E_D(x,y) \Bigr] = 
x e(x) -\frac{m}{M} f_1(x),\\ 
& & \int dy \,  \Bigl[2 \, {\rm Im} \, E_D(x,y) \Bigr] = 
i x h(x)
\label{Imeom3}.
\ea 
From this (and $iD^{\alpha}= i\partial^{\alpha} + g A^{\alpha}$) we see that 
the (T-odd) imaginary parts of the 
two-argument functions are related to the T-odd one-argument 
functions, as one expects.
So if time-reversal invariance is imposed, the imaginary parts of the e.o.m.\ 
Eqs.\ (\ref{Imeom1}), (\ref{Imeom2})
and (\ref{Imeom3}) become three trivial equalities. We like to point out that
if one integrates Eqs.\ (\ref{eom}) and (\ref{Imeom1}) over $x$, weighted with 
some test-function $\sigma(x)$, one arrives at the sum rules discussed in 
\cite{Efremov-Teryaev-84,Teryaev-97}.  

\section{Gluonic poles and time-reversal odd behavior}
\vspace{1mm}
\noindent
We are interested in the behavior of the quark-gluon correlation 
function $\Phi_A^\alpha$ 
in case $x=y$, when the gluon has
zero-momentum. For this purpose, we define ($\alpha$ is a transverse index)
\beq
\Phi_{Fij}^\alpha(x,y) \equiv \int \frac{d \lambda}{2\pi} 
\frac{d \eta}{2\pi} e^{i\lambda x} e^{i\eta (y-x)} 
\amp{P,S|\psibar_j (0) F^{+\alpha} (\eta n_-) 
\psi_i(\lambda n_-)| P,S}
\eeq
and $F^{\rho\sigma}(z)=\frac{i}{g} \left[ D^\rho(z), D^\sigma(z) \right]$. 
This matrix element has the same hermiticity,
but the opposite time-reversal 
behavior as $\Phi_A^\alpha$,
\ba
\left[\Phi_F^\alpha(x,y)\right]^\ast = 
- \gamma_5 C \,\Phi_{F\alpha}(x,y)\, 
C^\dagger \gamma_5 & &\quad \mbox{[Time\ reversal]}
\ea  
and we will parametrize it identically
with help of functions called $G_F(x,y), \tilde{G}_F(x,y), H_F(x,y)$ and
$E_F(x,y)$, noting that time-reversal implies 
that $G_F$ and $E_F$ are symmetric and 
thus may survive at $x=y$ (in contrast to $G_A(x,x)$ and $E_A(x,x)$).
In the gauge $A^+=0$ one has $F^{+\alpha}=\partial^+ A_T^\alpha$ and one 
finds by partial integration 
\beq
(x-y)\Phi_A^\alpha (x,y)= -i \Phi_F^\alpha(x,y).
\label{AvsF} 
\eeq 
If a specific Dirac projection of $\Phi_F^\alpha(x,x)$ is nonvanishing, then 
the corresponding projection of $\Phi_A^\alpha(x,x)$ has a pole, hence the 
name gluonic pole. An example  
is the function $T(x,S_T)\equiv \pi \text{Tr} \left[\Phi_F^\alpha(x,x)\, 
\epsilon_{T \beta \alpha} 
S_T^\beta \slsh{n_-} \right]/P^+ 
= 2\pi i M S_T^2 G_F(x,x)$ discussed by Qiu and 
Sterman in 
Ref.\ \cite{QS-91b}. 

In order to define Eq.\ (\ref{AvsF}) at the pole, one needs a
prescription, which is related to the choice of boundary conditions on 
$A_T^\alpha(\eta = \pm \infty)$ inside matrix elements. 
Possible inversions of $F^{+\alpha}$ =
$\partial^+A_T^\alpha$ are:
\ba
A_T^\alpha(\eta n_-) &=& A_T^\alpha(\infty) 
- \int_{-\infty}^{\infty} dz^-\ \theta(z^--\eta n_-)\,F^{+\alpha}(z^-)
\nonumber \\[3 mm]
&=& A_T^\alpha(-\infty) 
+ \int_{-\infty}^{\infty} dz^-\ \theta(\eta n_--z^-)\,F^{+\alpha}(z^-)
\nonumber \\[3 mm]
&=& \frac{A_T^\alpha(\infty) + A_T^\alpha(-\infty)}{2}
- \frac{1}{2}\int_{-\infty}^{\infty} dz^-\ \epsilon(z^--\eta n_-)
\,F^{+\alpha}(z^-).
\ea
One can use the representations for the $\theta$ and $\epsilon$ functions,
\beq
\pm i\theta(\pm x) = \int \frac{dk}{2\pi}\,\frac{e^{ikx}}{k\mp i\epsilon},
\qquad
i\epsilon(x) = \int \frac{dk}{2\pi}\,{\rm P}\frac{e^{ikx}}{k},
\eeq
to obtain 
\ba
\Phi^{\alpha}_A (x,y) &=& \delta(x-y)
\,\Phi^\alpha_{A(\infty)}(x)
+ \frac{-i}{x-y+i\epsilon}\,\Phi_F^\alpha(x,y)
\label{ieps}\\[3 mm]
&=&  
\delta(x-y)
\,\Phi^\alpha_{A(-\infty)}(x)
+ \frac{-i}{x-y-i\epsilon}\,\Phi_F^\alpha(x,y)
\nonumber \\[3 mm]
&=& 
\delta(x-y) 
\,\frac{\Phi^\alpha_{A(\infty)}(x) + \Phi^\alpha_{A(-\infty)}(x)}{2}
+ {\rm P}\frac{-i}{x-y}\,\Phi_F^\alpha(x,y),
\label{AvsFplusb}
\ea
where 
\beq
\delta(x-y)\,\Phi^\alpha_{A(\pm \infty)\,ij}
(x)
\equiv \int \frac{d \lambda}{2\pi} 
\frac{d \eta}{2\pi} e^{i\lambda x} e^{i\eta (y-x)} 
\amp{P,S|\psibar_j (0) gA_T^\alpha(\eta = \pm \infty) 
\psi_i(\lambda n_-)| P,S}.
\eeq
So Eq.\ (\ref{AvsFplusb}) shows the importance of boundary conditions
in the inversion of Eq.\ (\ref{AvsF}), if matrix elements containing
$A_T^\alpha(\eta = \pm \infty)$ do not vanish. When such matrix elements
vanish (implicitly assumed in \cite{Tangerman-Mulders-95a})
the pole prescription does not matter. 
Also one obtains 
\beq
2 \pi \, \Phi_F^\alpha(x,x) = 
\left[\Phi^\alpha_{A(\infty)}(x) - \Phi^\alpha_{A(-\infty)}(x)\right],
\eeq
which shows the
relation between the zero-momentum quark-gluon correlation function and the
boundary conditions. 

The behavior of $\Phi^\alpha_{A(\pm \infty)}(x)$ under time-reversal is:
\beq
\Phi^{\alpha \ast}_{A(\pm \infty)}(x)= \gamma_5 C
\,\Phi_{A(\mp \infty) \alpha}(x) \, C^\dagger \gamma_5.
\eeq
This relation implies that time-reversal invariance {\em only\/} allows for 
symmetric or antisymmetric boundary conditions. Both situations (if
nonvanishing) lead to a singularity in $\Phi_A^\alpha(x,y)$ at the point
$x=y$, but only the antisymmetric case will be called a gluonic pole.
The delta-function singularity in the case of nonvanishing symmetric 
boundary conditions will contribute to the functions 
$\tilde{G}_A(x,x)$ and $H_A(x,x)$ and hence, to T-even DFs. This 
would only affect the magnitude of double spin 
asymmetries. This case is also less interesting, because 
$\Phi_F^\alpha(x,x)=0$. 

We like to point out that so-called fermionic poles play a
role in off-forward scattering, such as prompt photon production 
\cite{ET-85,QS-91b,Korotkiyan-T-94,E-Korotkiyan-T-95}. Here a gluonic 
pole gives rise to an asymmetry 
proportional to $T(x,S_T)g(\bar x)$ (see Fig.\ \ref{prompt}). 
Fermionic poles do not contribute in case
of DY to this order.
\begin{figure}[htb]
\begin{center}
\leavevmode \epsfxsize=6.5cm \epsfbox{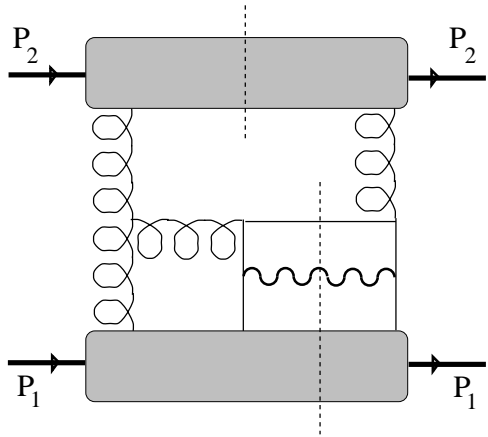}\\
\vspace{0.2 cm}
\caption{\label{prompt}A diagram
yielding a single transverse spin asymmetry in 
prompt photon production}
\end{center}
\end{figure}

\section{Effective T-odd distribution functions}
\vspace{1mm}
\noindent
To study the effect of gluonic poles 
we consider the (nonvanishing) antisymmetric boundary 
condition $\Phi^\alpha_{A(\infty)}(x)= 
- \Phi^\alpha_{A(- \infty)}(x)$, which implies 
\ba
&& \pi\, \Phi_F^\alpha(x,x) =
\Phi^\alpha_{A(\infty)}(x),\\
&& \Phi^{\alpha \ast}_{A(\pm \infty)}(x)= - \gamma_5 C
\,\Phi_{A(\pm \infty) \alpha}(x) \, C^\dagger \gamma_5.
\ea
In the calculation of the cross-section
one always encounters the pole of the matrix element (in this case in the 
principal value prescription) multiplied with the propagator in the 
hard subprocess (having a causal prescription),
\ba
\Phi_A^{\alpha \, \text{eff}} (y,x) & \equiv & 
\frac{x-y}{x-y+i\epsilon}\,\Phi_A^\alpha(y,x)\\
&=& \frac{-i}{x-y+i\epsilon}\,\Phi_F^\alpha(y,x)\\[2 mm]
&=& \Phi_A^\alpha(y,x) - \pi\,\delta(x-y) \,\Phi_F^\alpha(y,x).
\ea
The time-reversal constraint applied to $\Phi_A^\alpha(x,y)$ implies 
the analogue of Eq.\ (\ref{TinvA}), while   
$\Phi_F^\alpha(x,y)$ has the opposite behavior under time-reversal compared to
$\Phi_A^\alpha(x,y)$. Thus for $\Phi_A^{\alpha \, \text{eff}} (x,y)$ one 
does not have definite behavior under time-reversal symmetry. 
Specifically, the allowed T-even functions of 
$\Phi_F^\alpha(x,x)$, $G_F(x,x)$ and $E_F(x,x)$, can be identified with T-odd 
functions in the effective 
correlation function $\Phi_A^{\alpha \, \text{eff}}$. 
This implies that $G_A^{\text{eff}}(x,y)$ and $E_A^{\text{eff}}(x,y)$ will 
have an imaginary part and this gives rise to two 
"effective" time-reversal-odd DFs 
via the imaginary part of the e.o.m. 

To say it again in a different way: 
by partial integration we find for instance 
\beq 
G_A(x,y)= \frac{-i}{x-y} G_F(x,y).
\label{GAvsGF}
\eeq
If one applies time-reversal invariance, $G_A(x,y)$ will be a real function
and $G_F(x,y)$ imaginary. So one expects the pole prescription to be the
principal value. But when convoluting the pole of the matrix element (with
the principal value prescription) with the propagator in the 
hard subprocess (with a causal prescription), 
it is formally possible to shift the imaginary 
part from the pole of the latter to the pole of the 
former. This will effectively give rise to a causal prescription in Eq.\
(\ref{GAvsGF}), instead of a principal value (but without the additional 
boundary term required by time-reversal, cf.\ Eq.\ (\ref{ieps})).
This implies that $G_A(x,y)$ (and also $E_A(x,y)$) will 
effectively acquire an imaginary part.  

For simplicity we neglect intrinsic tranverse momentum, 
thus we assume $\Phi_{A(\infty)}^\alpha(x) = 
\Phi_{D(\infty)}^\alpha(x)$. 

So, by identification we have 
\ba
i \pi \, G_F(x,x)& =&  \int dy \,  
\text{Im} \, G_A^{\text{eff}}(y,x),\\
i \pi \, E_F(x,x) &= & \int dy \, 
\text{Im} \, E_A^{\text{eff}}(y,x)
\ea
and then it follows from the e.o.m.\ that  
\ba
x f_T^{\text{eff}}(x) & =& i \pi G_F(x,x) = 
\frac{1}{2 M S_T^2} T(x,S_T), \\
x h^{\text{eff}}(x) 
&= &2 i \pi E_F(x,x) = \frac{- i \pi}{2MP^+}
\text{Tr}\left[ \Phi_F^\alpha (x,x)\, \gamma_{T \alpha} \slsh{n_-}\right].
\label{htildeproj}
\ea 
The function
$e_L^{\text{eff}}$ receives no gluonic pole contribution, since 
time-reversal symmetry requires $H_F(x,x)=0$. 

Of course, the mechanism for generating finite projections of 
$\Phi_F^\rho(x,x)$ remains unknown. We just can conclude that if
there is indeed a non-zero gluonic pole (in the case of 
non-zero antisymmetric boundary conditions), then at twist-three there are two
non-zero ``effective'' T-odd DFs, 
namely $f_T$ and $h$. 
The first one generates the twist-three single spin asymmetry found by Hammon\ 
{\em et al.\/}~\cite{Hammon-97}, 
in their notation it is proportional to $T(x,x)$. The second one leads to a
new asymmetry (see next section). 
Summarizing, we find for the T-{\em even\/} parametrization of
$\Phi^\alpha_{A(\infty)}(x)$, 
\ba
\Phi^\alpha_{A(\infty)}(x) &=& 
- \frac{i x M}{2}\bigg[ f_T^{\text{eff}}(x) 
\,i\epsilon_T^{\alpha\beta} 
S_{T \, \beta} \mbox{$\not\! P \,$} + 
\frac{1}{2} h^{\text{eff}}(x)
\gamma_T^\alpha \mbox{$\not\! P \,$} \bigg].
\label{paramPhiAinf}
\ea

The antisymmetric nonvanishing boundary condition for 
$\Phi^\alpha_{A(\pm\infty)}(x)$ might arise from a linear A-field, 
giving a constant field strength (cf.\ e.g.\ \cite{Schaefer-93,Ehrnsperger}). 
One might also think of an instanton background field. In both cases one 
should interpret infinity to mean 'outside the proton radius'. 
Also, the constant field strength should be understood as an average value
of the gluonic chromomagnetic field, which is non-zero due to a correlation
with the direction of the proton spin.
The large distance origin of the asymmetries arising from such a gluonic 
pole is apparent. 

\section{Single spin asymmetries in the Drell-Yan process}
\vspace{1mm}
\noindent
We will now discuss the single spin asymmetries in the 
Drell-Yan process in case one integrates over
transverse photon momentum. So one uses the above parametrizations of the
correlation functions in the
expression for the integrated hadron tensor, 
which after contraction with the 
lepton tensor yields the cross-section. 

\begin{figure}[htb]
\begin{center}
\leavevmode \epsfxsize=10cm \epsfbox{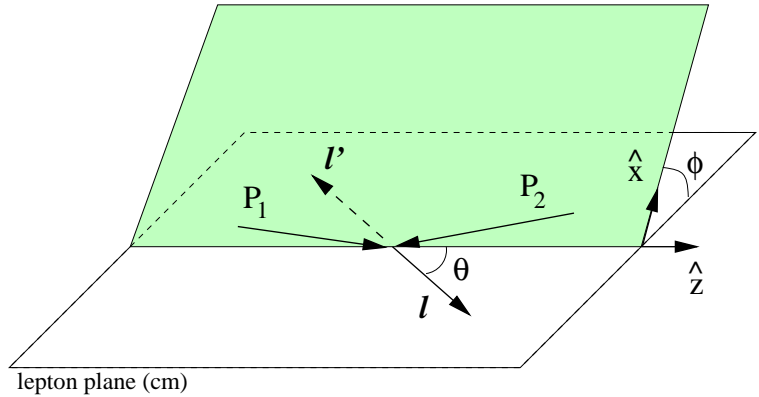}
\vspace{0.2 cm}
\caption{\label{DYkin} Kinematics of the Drell-Yan process in
the lepton center of mass frame}
\end{center}
\end{figure}
Under the assumption that $\Phi_{A(\infty)}^\alpha= 
\Phi_{D(\infty)}^\alpha$ we find the following single spin asymmetry 
(hadron-two unpolarized), given in the lepton center of mass frame:
\ba
\lefteqn{A_T = \frac{4\sin(2\theta)\;\sin(\phi_{S_1})}{1 + \cos^2\theta} 
\frac{|\bm S_{1T}^{}|}{Q}} \nonumber \\
&& \quad \mbox{} \times \sum_{a}e_a^2
\;\bigg[M_1 \, x \,f_T^a (x)
f_1^{\bar a}(\bar x) + M_2\, h_1^a(x) \bar x \, 
h^{\bar a}(\bar x) 
\bigg]\Bigg/ \sum_{a}e_a^2 \; f_1^a (x) f_1^{\bar a} (\bar x),
\ea
where $\phi_{S_1}$ is the angle between $S_{1T}^{}$ and the perpendicular part
of the lepton
momentum $l$, $\hat l_\perp^\mu \equiv \left( g^{\mu \nu}-\hat t^{\{ 
\mu} \hat t^{\nu \} } + \hat z^{\{ 
\mu} \hat z^{\nu \} } \right) l_\nu$.
The first term in the asymmetry (proportional to $f_T$) is 
the one discussed in
\cite{Hammon-97} (in their notation it is proportional to $T(x,x) q(y)$),
which will also be present in DIS ($f_1(\bar x) = \delta(1-\bar x)$).
The second term is 
another, new single spin asymmetry arising in DY from a gluonic pole. It
is not proportional to $T(x,S_T)$, but to a chiral-odd projection of
$\Phi_F^\alpha$ in the point $x=y$, cf.\ Eq.\ (\ref{htildeproj}). 

\section{Conclusions}
\vspace{1mm}
\noindent
We have demonstrated for the Drell-Yan process that 
the effects of so-called gluonic poles in twist-three hadronic matrix 
elements cannot be distinguished from those of T-odd distribution
functions. Imaginary phases arising from hard subprocesses together 
with gluonic poles
give rise to {\em effective\/} T-odd distribution functions. 
This leads to single spin asymmetries for the Drell-Yan process, such as the 
one found recently by Hammon {\em et al.\/}~\cite{Hammon-97}. We have found
a similar asymmetry arising from a gluonic pole, which involves chiral 
odd distribution functions. 
We have moreover shown that the presence of gluonic poles is in accordance 
with time-reversal invariance and requires large distance gluonic fields with
antisymmetric boundary conditions. 

\vspace{1cm}
\noindent 
We thank A. Sch\"afer for useful discussions. 
This work was in part supported by the Foundation for 
Fundamental 
Research on Matter
(FOM) and the National Organization for Scientific Research (NWO). 
It is also performed in the framework of Grant
96-02-17631 of the Russian Foundation for Fundamental Research
and Grant $N^o_-$ 93-1180 from INTAS.

\end{document}